%% file: main.tex
\documentclass{article} 
\usepackage{iclr2026_conference,times}

\input{math_commands.tex}

\usepackage{hyperref}
\usepackage{url}
\usepackage{graphicx}
\usepackage{booktabs}

\title{ATTNSOM: Learning Cross-Isoform Attention for Cytochrome P450 Site-of-Metabolism}

\author{Hajung Kim, Eunha Lee, Sohyun Chung, Jueon Park, Suengheun Baek \& Jaewoo Kang \thanks{ Corresponding author.} \\
Department of Computer Science and Engineering\\
Korea University\\
Seoul, 02841, South Korea \\
\texttt{kangj@korea.ac.kr} \\
}

\begin{document}

\maketitle

\begin{abstract}
Identifying metabolic sites where cytochrome P450 enzymes metabolize small-molecule drugs is essential for drug discovery. Although existing computational approaches have been proposed for site-of-metabolism prediction, they typically ignore cytochrome P450 isoform identity or model isoforms independently, thereby failing to fully capture inherent cross-isoform metabolic patterns. In addition, prior evaluations often rely on top-k metrics, where false positive atoms may be included among the top predictions, underscoring the need for complementary metrics that more directly assess binary atom-level discrimination under severe class imbalance. We propose ATTNSOM, an atom-level site-of-metabolism prediction framework that integrates intrinsic molecular reactivity with cross-isoform relationships. The model combines a shared graph encoder, molecule-conditioned atom representations, and a cross-attention mechanism to capture correlated metabolic patterns across cytochrome P450 isoforms. The model is evaluated on two benchmark datasets annotated with site-of-metabolism labels at atom resolution. Across these benchmarks, the model achieves consistently strong top-k performance across multiple cytochrome P450 isoforms. Relative to ablated variants, the model yields higher Matthews correlation coefficient, indicating improved discrimination of true metabolic sites. These results support the importance of explicitly modeling cross-isoform relationships for site-of-metabolism prediction. The code and datasets are available at \url{https://github.com/dmis-lab/ATTNSOM}.
\end{abstract}

\input{1_introduction}

\input{2_method}

\input{3_results}

\input{4_analysis}

\input{5_conclusion}

\input{6_extra}

\bibliography{iclr2026_conference}
\bibliographystyle{iclr2026_conference}

\end{document}

%% file: math_commands.tex
\usepackage{amsmath,amsfonts,bm}

\def\eqref#1{equation~\ref{#1}}

\def\1{\bm{1}}

\DeclareMathAlphabet{\mathsfit}{\encodingdefault}{\sfdefault}{m}{sl}
\SetMathAlphabet{\mathsfit}{bold}{\encodingdefault}{\sfdefault}{bx}{n}



%% file: 1_introduction.tex
\section{Introduction}
\label{sec:intro}

Xenobiotics are exogenous chemical compounds not synthesized endogenously in living organisms. These compounds, including drugs, undergo enzymatic metabolism and are converted into more hydrophilic derivatives to facilitate elimination~\citep{guengerich2008cytochrome}. This process occurs primarily in the hepatic system and comprises two phases: Phase I and Phase II. Phase I reactions include oxidation, reduction, and hydrolysis, introducing polar functional groups into compounds. These reactions are predominantly catalyzed by cytochrome P450 (CYP) enzymes~\citep{guengerich2006cytochrome}. The resulting metabolites may undergo Phase II metabolism, in which transferases mediate conjugation with endogenous polar moieties, such as glucuronosyltransferases, sulfotransferases, and glutathione S-transferases, enabling efficient excretion. CYP enzymes are involved in over 90\% of drug metabolism and represent the most prominent drug-metabolizing enzyme family in humans~\citep{rendic2015survey}. Among the human CYP superfamily, CYP1, CYP2, and CYP3 families account for the metabolism of around 80\% of approved drugs~\citep{ingelman2004human}.

Although metabolic clearance is necessary for the elimination of xenobiotics and detoxification, excessive CYP-mediated metabolism often results in rapid drug clearance, leading to a shorter half-life. This, in turn, can compromise the maintenance of therapeutically relevant drug concentrations, ultimately limiting therapeutic efficacy~\citep{wang2019induction}. To preserve therapeutic efficacy by extending metabolic stability, medicinal chemists employ strategic structural modifications that substitute metabolically labile sites with alternative molecular fragments~\citep{smith2017relevance}. A representative example of drug optimization driven by pharmacokinetic considerations is deutivacaftor, an FDA-approved Cystic Fibrosis Transmembrane conductance Regulator (CFTR) modulator that is a deuterated analogue of ivacaftor. Ivacaftor is primarily oxidized by CYP3A4 at a single tert-butyl group, leading to metabolic inactivation, a short half-life, and the need for twice daily administration. Targeted deuterium substitution at this metabolically labile position suppressed the dominant metabolic pathway and improved pharmacokinetic stability, allowing once daily dosing~\citep{di2023deuterium,mullard2025fda}. The identification and characterization of CYP-mediated site-of-metabolism (SOM) represent a critical step in drug optimization and development.

To enable the identification of CYP-mediated SOMs, substantial efforts have focused on experimentally characterizing CYP-catalyzed biotransformation mechanisms and annotating metabolic reactivity at atomic level resolution. These advances have facilitated the development of computational methods for predicting SOMs. Over the past decade, SOM prediction has progressed from rule-based chemical heuristics methods. Early computational predictors were predominantly grounded in predefined chemical rules. SMARTCyp~\citep{rydberg2010smartcyp} ranks candidate metabolic sites using rule-based scores derived from intrinsic reactivity and steric accessibility. RS-Predictor~\citep{zaretzki2011rs} utilizes atom-level topological and quantum chemical descriptors that are optimized within a multiple instance learning framework, MIRank~\citep{bergeron2008multiple}, to preferentially rank reactive over non-reactive ones.

With the accumulation of larger volumes of metabolic data, neural network models were introduced to capture more complex relationships between molecular structure and metabolic reactivity. XenoSite~\citep{zaretzki2013xenosite} represents an early example of this approach by combining atom-level and molecule-level descriptors within a multilayer perceptron to predict CYP-mediated SOMs. FAME 3~\citep{sicho2019fame}, which is based on extremely randomized trees, further improved predictive performance by leveraging circular fingerprints centered on the atom. GLMCyp~\citep{huang2025glmcyp} uses multi-dimensional molecular representations derived from Uni-Mol~\citep{zhou2023uni} and protein embeddings obtained from ESM2~\citep{lin2023evolutionary}, which are integrated via attention mechanisms to capture cross-modal dependencies.

Despite advances in CYP-mediated SOM prediction, several important limitations remain. With the exception of GLMCyp, most existing methods do not explicitly incorporate CYP isoform information, but instead learn generic associations between atom-level descriptors and SOM labels. Consequently, CYP-dependent metabolic preferences and relationships across isoforms remain under-exploited. Experimental evidence indicates that human CYP isoforms exhibit broad and partially overlapping substrate specificity, resulting in structurally the same compounds can be metabolized by multiple CYP isoforms~\citep{mustafa2019differing, yan2025elucidating}. This observation implies that CYP-mediated metabolism has correlated patterns of metabolic reactivity across isoforms.

\begin{figure}[t]
  \centering
  \includegraphics[width=0.8\textwidth, alt={Heatmap displaying pairwise similarity scores of SOM patterns among nine CYP isoforms, with numerical values shown in each cell and a hierarchical clustering dendrogram above the matrix grouping isoforms based on similarity.}]{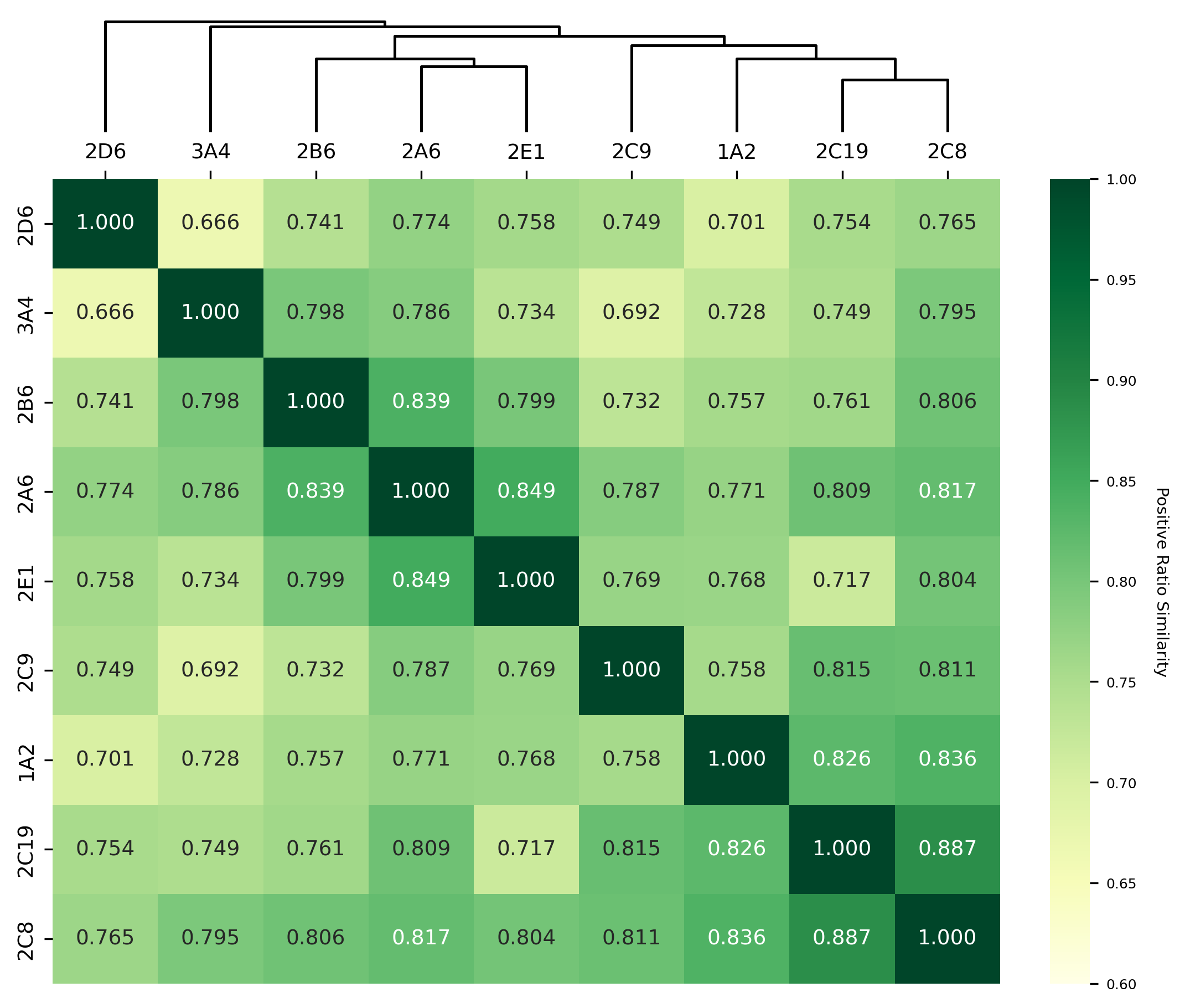}
  \caption{Hierarchical clustering and heatmap of pairwise SOM pattern similarity across CYP isoforms in the Zaretzki dataset. The dendrogram illustrates the relational structure among isoforms based on shared metabolic patterns.}

  \label{fig:heatmap}
\end{figure}

To empirically analyze these cross-isoform relationships, we examine pairwise Jaccard similarities in metabolic sites across CYP isoforms. Figure~\ref{fig:heatmap} presents the resulting pairwise similarities based on substrate molecules annotated in the publicly available dataset released by the XenoSite study~\citep{zaretzki2013xenosite}. Similarity values represent the extent to which annotated metabolic sites overlap between two isoforms. The analysis reveals relatively high similarity scores across isoform pairs, ranging from a minimum of 0.666 to a maximum of 0.887, indicating substantial overlap in SOM patterns across multiple CYP isoforms. These observations are consistent with following properties. The CYP enzymes share a common catalytic architecture as heme iron–dependent mono-oxygenases and operate through a conserved oxidative mechanism. As a consequence, the repertoire of feasible reactions is restricted to a limited set of processes, including hydroxylation, dealkylation, epoxidation, and heteroatom oxidation~\citep{guengerich2008cytochrome, de2007cytochrome}. This constraint confines metabolic reactivity to a small number of chemically favorable positions within a molecule, thereby leading to overlapping SOM patterns across CYP isoforms.

Beyond the need to incorporate CYP isoform information, evaluation practices used in prior studies also lack metrics suitable for assessing atom-level discrimination under severe class imbalance. Metrics that are commonly reported, such as Top-2 accuracy, focus on only a small number of highest ranked candidate sites and provide limited information about a model’s ability to distinguish true reactive from non-reactive atoms. 

To address these limitations, we introduce a unified framework composed of the following components:
\begin{itemize}
    \item \textbf{Shared graph encoder for intrinsic chemical reactivity.}  
    A graph based encoder is shared to learn atom-level representations that capture intrinsic chemical reactivity independent of enzyme context.

    \item \textbf{Molecule conditioned atom representations.}  
    Feature wise linear modulation (FiLM) is applied to atom representations to incorporate molecule specific context, allowing atom-level features to vary across different molecular structures.

    \item \textbf{Modeling Shared Metabolic Patterns Across CYP Isoforms.}  
    A cross attention mechanism between atom representations and CYP isoform embeddings is used to capture relationships among CYP isoforms. Instead of treating isoforms independently, the model exploits correlated metabolic patterns across multiple isoforms to help guide SOM prediction.

    \item \textbf{Complementary metrics for binary discrimination.} We use the Matthews correlation coefficient (MCC)~\citep{chicco2020advantages} to capture overall binary classification.
\end{itemize}
The proposed model achieves higher average Top-2 performance and consistently improved MCC across all CYP isoforms, indicating more reliable atom-level discrimination between reactive and non-reactive sites. Additional analyses further demonstrate strong generalization and the ability to capture cross-isoform relationships.

%% file: 2_method.tex
\section{Materials and Methods}

\subsection{Dataset}

In this study, we use a publicly available site-of-metabolism (SOM) dataset released by the XenoSite. Hereafter, we refer to this dataset as the \emph{Zaretzki dataset}. The dataset contains 679 substrate molecules with experimentally annotated SOMs across nine major human cytochrome P450 isoforms: CYP1A2, CYP2A6, CYP2B6, CYP2C8, CYP2C9, CYP2C19, CYP2D6, CYP2E1, and CYP3A4, resulting in a total of 2,003 molecule-isoform instances. Table~\ref{tab:cyp_substrates} shows that substrate counts vary across isoforms. Each substrate is annotated for multiple CYP isoforms, with an average of approximately three per molecule. Across all isoforms, molecules contain about 22 atoms on average, whereas only 1.42 atoms per molecule are annotated as metabolic sites. At the atom-level, metabolized atoms represent a small fraction of all atoms, resulting in a strongly imbalanced prediction setting.

In addition to the Zaretzki dataset, we also use a SOM dataset released by AstraZeneca~\citep{chen2025metabolite}. Hereafter, we refer to this dataset as the \emph{AZ-ExactSOM}. This dataset consists of 120 publicly available compounds annotated with high confidence, atom resolved SOMs obtained from human hepatocyte assays. In this study, we consider only exact site annotations, corresponding to metabolically labile atoms that can be directly identified from experimentally resolved metabolite structures. The AZ-ExactSOM dataset does not include CYP isoform specific labels. Instead, annotated sites represent aggregate metabolic outcomes observed in human hepatocytes. Across the AstraZeneca compound set, molecules contain an average of 2.3 exact SOMs per compound. Similar to the Zaretzki dataset, only a small proportion of atoms are labeled as metabolic sites, resulting in a substantial imbalance between metabolized and non metabolized atoms at the atom level.

\input{tables/dataset}

\subsection{Model}

We propose ATTNSOM, a CYP-aware framework for SOM prediction that predicts atom-level metabolic likelihoods for a given molecular graph and target CYP isoform by integrating intrinsic chemical reactivity with molecule-conditioned and cross-isoform information. The model architecture comprises three components. A shared graph encoder first learns atom level representations that capture intrinsic reactivity patterns from molecular structure. Second, a graph-level feature-wise linear modulation component conditions these atom representations on global molecular context, transforming the atom feature space to vary across molecules. Third, the cross-attention module between atoms and all CYP isoforms captures cross-isoform relationships, allowing shared metabolic patterns to help predict SOMs.

\paragraph{\textbf{Shared graph encoder}}
Each molecule is represented as a graph \( G = (V, E) \), where nodes (\(V\)) correspond to atoms and edges (\(E\)) represent chemical bonds. We adopt GraphCliff~\citep{kim2025graphcliff}, which jointly models short-range local interactions and long-range structural context through multi-layer message passing. In conventional deep graph encoders, increasing message-passing depth often leads to oversmoothing, reducing the discriminability of atom-level representations. GraphCliff addresses this limitation by explicitly gating the flow of short- and long-range information, thereby preserving local atom-level features relevant to the target property. This design is well suited for SOM prediction, where evaluation is performed at the atom level and accurate modeling of metabolism-related atom-level representations is critical. The graph encoder is shared across all CYP isoforms to capture isoform-agnostic molecular reactivity, providing a common foundation for subsequent modeling of isoform-dependent effects. Given a molecular graph with \( N = |V| \) atoms, the shared encoder produces atom-level representations \( \{\mathbf{n}_i\}_{i=1}^{N} \) and a graph-level representation \( \mathbf{g} \), where \( \mathbf{n}_i \in \mathbb{R}^H \) denotes the embedding of atom \( i \), and \( \mathbf{g} \in \mathbb{R}^H \) summarizes the global molecular context.

\begin{figure*}[htbp]
  \centering
  \includegraphics[width=1\textwidth, alt={Schematic diagram of the ATTNSOM model architecture showing a graph-based molecular encoder for substrates, CYP isoform embeddings, an attention head combining atom and isoform representations, and a prediction head producing atom-level sites of metabolism.}]{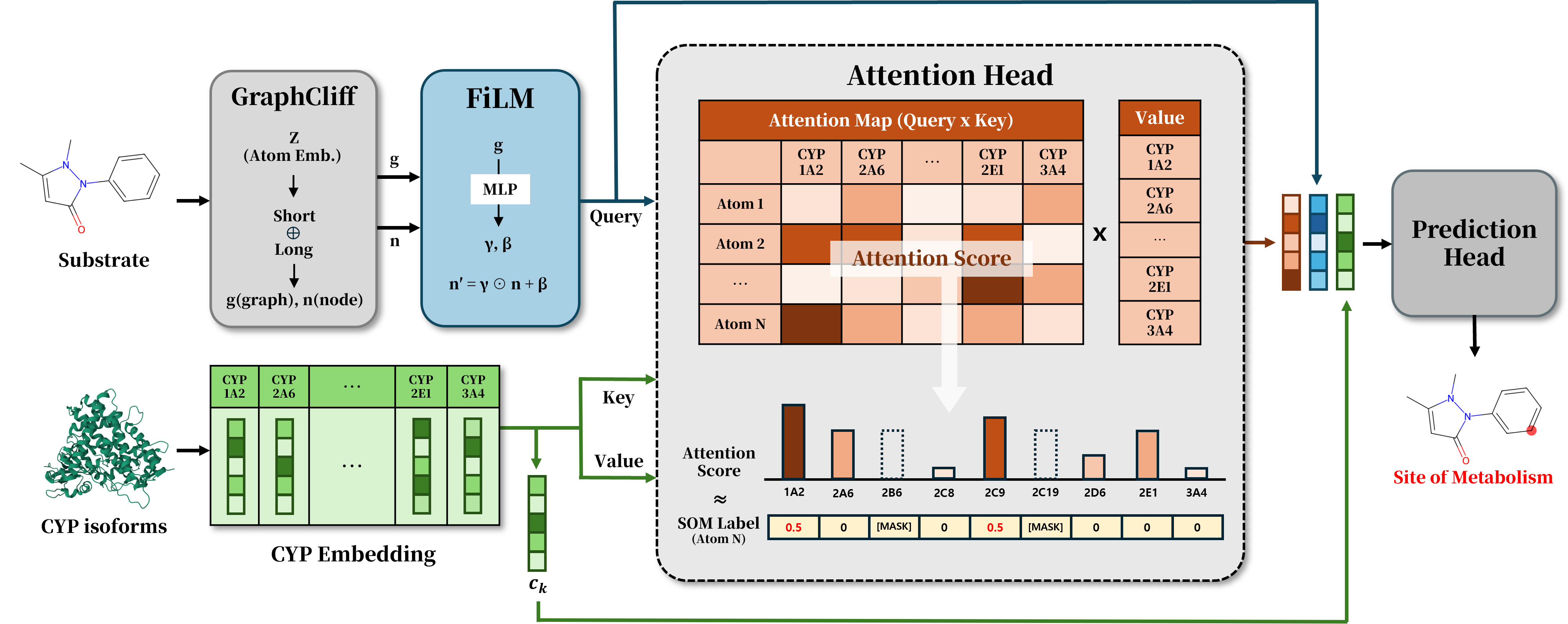}
  \caption{Overview of the ATTNSOM architecture for atom-level SOM prediction across CYP isoforms. The framework combines a shared graph-based molecular encoder with molecule-conditioned atom representations and a cross-attention module that models inter-isoform metabolic relationships. By integrating atom-level features with relational information across CYP isoforms, ATTNSOM captures both shared and distinct metabolic patterns.}
  \vspace{-1em}
  \label{fig:architecture}
\end{figure*}

\paragraph{\textbf{Molecule-conditioned atom representations via FiLM}}
A shared graph encoder can capture informative atom-level representations. However, atom reactivity varies across molecules due to molecule-specific context and positional effects. To explicitly condition atom representations on their molecular context and distinguish such cases, we employ feature-wise linear modulation (FiLM). Specifically, the graph-level representation \( \mathbf{g} \) is transformed into FiLM parameters through a multilayer perceptron:
\begin{equation}
(\boldsymbol{\gamma}, \boldsymbol{\beta}) = f_{\text{FiLM}}(\mathbf{g}),
\end{equation}
where \( \boldsymbol{\gamma}, \boldsymbol{\beta} \in \mathbb{R}^H \).
Each atom representation is then modulated as
\begin{equation}
\mathbf{n}_i' = (1 + \tanh(\boldsymbol{\gamma})) \odot \mathbf{n}_i + \boldsymbol{\beta},
\end{equation}
where \( \odot \) denotes element-wise multiplication. This formulation enables atom representations to adapt to molecule-specific context while preserving the atom-level reactivity patterns.

\paragraph{\textbf{Cross-attention between atoms and CYP isoforms}}
To explicitly model relationships among CYP isoforms and incorporate cross-isoform metabolic information into SOM prediction, we introduce a cross-attention mechanism. Each CYP isoform is associated with a learnable embedding \( \mathbf{c}_k \in \mathbb{R}^H \), where \( k \in \{1, \dots, K\} \) indexes the set of CYP isoforms. Given molecule-conditioned atom representations \( \mathbf{n}_i' \), attention scores between atom \( i \) and all CYP isoforms are computed as
\begin{equation}
\label{eq:cross_attention}
\alpha_{ik} = \frac{\exp \left( \frac{ (\mathbf{W}_q \mathbf{n}_i')^\top (\mathbf{W}_k \mathbf{c}_k) }{\sqrt{d}} \right)}
{\sum_{k'} \exp \left( \frac{ (\mathbf{W}_q \mathbf{n}_i')^\top (\mathbf{W}_k \mathbf{c}_{k'}) }{\sqrt{d}} \right)},
\end{equation}
where \( \mathbf{W}_q \) and \( \mathbf{W}_k \) are learnable projection matrices, \( d \) denotes the attention head dimension, and the summation over \( k' \in \{1, \dots, K\} \) ensures softmax normalization across all CYP isoforms. The attended representation ($\mathbf{n}_i^{\text{attn}}$) for atom \( i \) is computed as an attention-weighted sum of value-projected CYP isoform embeddings,
\begin{equation}
\mathbf{n}_i^{\text{attn}} = \sum_{k=1}^{K} \alpha_{ik} \mathbf{W}_v \mathbf{c}_k,
\end{equation}
where \( \mathbf{W}_v \) is a learnable value projection matrix. We adopt an atom centric attention formulation in which atom representations serve as queries, while CYP isoform embeddings are used as keys and values. From this perspective, each atom effectively queries which CYP isoforms are most relevant to its metabolic reactivity. This design choice aligns naturally with the objective of SOM prediction, which seeks to determine whether a given atom constitutes a potential metabolic site. By allowing each atom to attend over multiple CYP isoforms simultaneously, the model can integrate shared and correlated metabolic patterns across related enzymes, thereby enabling isoform-aware atom-level prediction. An inverse configuration that treats CYP embeddings as queries, as used in GLMCyp, is effective for modeling enzyme specificity. However, this formulation is less directly aligned with atom-level discrimination. Our atom centric design more explicitly reflects the prediction objective and naturally supports cross-isoform relationships.

\paragraph{\textbf{Prediction Head}}
For a given target CYP isoform \( t \), SOM prediction is performed by integrating atom-level, cross-isoform relational, and target CYP isoform information. Specifically, the FiLM-modulated atom representation \( \mathbf{n}_i' \), the attended atom representation \( \mathbf{n}_i^{\text{attn}} \), and the target CYP isoform embedding \( \mathbf{c}_t \) are concatenated as
\begin{equation}
\mathbf{z}_{i} = [\mathbf{n}_i' \oplus \mathbf{n}_i^{\text{attn}} \oplus \mathbf{c}_t] ,
\end{equation}
and passed through a multilayer perceptron to produce an atom-wise logit for CYP isoform \( t \).

\subsection{Training Objective}
The training objective consists of a primary atom-level classification loss and an auxiliary attention alignment loss:
\begin{equation}
\mathcal{L} = \mathcal{L}_{\text{main}} + \lambda_{\text{attn}} \mathcal{L}_{\text{attn}},
\end{equation}
where $\lambda_{\text{attn}}$ controls the contribution of the auxiliary objective. For the main task of atom-level SOM prediction, we supervise the predicted atom-wise logits using a Focal loss, which mitigates the severe class imbalance between reactive and non-reactive atoms by down-weighting easy negatives and emphasizing hard, informative examples. To encourage biologically meaningful cross-isoform interactions, we introduce an auxiliary attention alignment loss that supervises the attention scores using experimentally annotated SOM information across CYP isoforms. For each molecule $m$, let ${c}_m \subseteq \{1,\dots,K\}$ denote the set of CYP isoforms for which experimental annotations are available. For each atom $i$ in molecule $m$, we define the set of positive isoforms $\mathcal{P}_i \subseteq {c}_m$ as the CYP isoforms for which the atom–CYP pair $(i, c)$ is annotated as positive. The remaining isoforms are treated as negative for atom $i$. We construct a soft target label by uniformly distributing probability mass over the  positive isoforms:
\begin{equation}
y_{ik}=
\begin{cases}
\frac{1}{|\mathcal{P}_i|}, & k \in \mathcal{P}_i,\\
0, & k \in \{\mathbf{c}_m - \mathcal{P}_i\},
\end{cases}
\end{equation}
such that an atom annotated for multiple isoforms receives fractional supervision (e.g., $1/3$ each when three isoforms are positive).

Let $\alpha_{ik}$ denote the attention weight produced by the cross-attention module (Eq.~\ref{eq:cross_attention}). The auxiliary loss is computed only on positive or negative atom-isoform annotations available in the training set and explicitly masks all experimentally unobserved pairs, including any atom-isoform combinations belonging to the test split. We define a binary mask
\begin{equation}
M_{ik}=
\begin{cases}
1, & k \in {c}_m,\\
0, & k \notin {c}_m,
\end{cases}
\end{equation}
which excludes isoforms that are not annotated for molecule $m$. Using this mask, the auxiliary attention alignment loss is defined as a masked soft-label binary cross-entropy:
\begin{equation}
\mathcal{L}_{\text{attn}}
=
\sum_{i}\sum_{k=1}^{K}
M_{ik}\,
\mathrm{BCE}\!\left(\alpha_{ik},\, y_{ik}\right).
\end{equation}
This masking strategy prevents penalization of unobserved atom-isoform interactions and ensures that the auxiliary supervision is applied only where experimental evidence exists. Overall, the auxiliary objective softly aligns the learned attention patterns with known cross-isoform metabolic associations while preserving flexibility for generalization beyond the observed annotation space.

\subsection{Implementation Details and Computational Cost}
Our model is trained using the Adam optimizer with a learning rate of \(1\times10^{-4}\) and weight decay of \(1\times10^{-4}\). The hidden dimension is set to 256, and the shared graph encoder consists of four message-passing layers. Training is performed with a batch size of 32 for up to 50 epochs using Focal Loss (\(\gamma=1\)) and an auxiliary attention loss weighted by \(\lambda_{\text{attn}}=0.5\). We adopt a 10-fold cross-validation protocol for training and evaluation. In each fold, the data are split into training and test sets, and 5\% of the training data is randomly held out as a validation set. Model checkpoints are selected based on the lowest validation loss. Under this training and evaluation setup, the proposed model exhibits favorable computational efficiency. The complete 10-fold cross-validation training, including validation-based checkpoint selection, completes in approximately 40 minutes end-to-end. At inference time, the model processes instances efficiently at approximately \(0.0004\) s per instance, substantially faster than existing methods (FAME 3: \(\sim15.95\) s per molecule; SMARTCyp: \(\sim1.5\) s per molecule). We note that both FAME 3 and SMARTCyp are accessed through external APIs, and their reported runtimes include network and service overhead. GPU-based training and inference require approximately 1500 MB of memory. Overall, these results indicate that ATTNSOM is computationally efficient and suited for large-scale screening.

%% file: tables/dataset.tex

\begin{table}[h]

\caption{Dataset statistics for Zaretzki dataset.}
\label{tab:cyp_substrates}
\centering
\scalebox{0.85}{
\begin{tabular}{l@{\hspace{4pt}}ccccccccc}
\toprule
 & 1A2 & 2A6 & 2B6 & 2C8 & 2C9 & 2C19 & 2D6 & 2E1 & 3A4 \\
\midrule
\# Substrates
& 271 & 105 & 151 & 142 & 226 & 218 & 270 & 145 & 475 \\

Avg. \# SoMs
& 1.5 & 1.5 & 1.4 & 1.4 & 1.4 & 1.4 & 1.4 & 1.5 & 1.4 \\

Avg. \# Atoms
& 19.7 & 15.5 & 18.9 & 21.7 & 21.1 & 21.0 & 20.9 & 15.5 & 25.1 \\
\bottomrule
\end{tabular}}
\end{table}


%% file: 3_results.tex
\section{Results}

We evaluate the proposed model on two site-of-metabolism (SOM) benchmarks. The Zaretzki dataset provides CYP isoform specific annotations and is evaluated using a 10 fold cross-validation with random splitting. Atom level classification performance is assessed using the Top-$k$ accuracy and Matthews correlation coefficient (MCC). Direct MCC comparison with XenoSite, SMARTCyp, and RS-Predictor is not feasible, as these methods are accessible only via web servers or are no longer available as services, and do not provide isoform-resolved atom-level scores. We therefore restrict comparisons with these methods to Top-$k$ metrics reported in prior work, while MCC-based comparisons are conducted exclusively among trainable baselines and ablated variants evaluated under identical splits. The models are also evaluated on the AZ-ExactSOM dataset. The two benchmarks evaluate a model's ability to both rank and discriminate reactive sites under severe class imbalance.

\input{tables/result_top2}

\paragraph{\textbf{Performance on the Zaretzki Dataset}}
Table~\ref{tab:top2} summarizes the Top-2 SOM prediction accuracy across nine CYP isoforms. ATTNSOM achieves the highest average Top-2 accuracy (0.871) among baselines. Across individual isoforms, ATTNSOM shows strong and stable performance, achieving the best Top-2 accuracy on six out of nine CYP isoforms, including CYP2A6, CYP2B6, CYP2C8, CYP2C9, CYP2C19, and CYP2E1. This performance indicates that ATTNSOM effectively captures cross-isoform SOM preferences, particularly for CYP2C9 (0.938) and CYP2C19 (0.922). In the XenoSite study, the evaluation is conducted using a leave-one-out approach. In reported results, it shows competitive performance across several CYP isoforms and attains the highest Top-2 accuracy for CYP1A2, CYP2D6, and CYP3A4, with an average score of 0.870. RS-Predictor is evaluated under a cross-validation protocol and yields lower Top-2 accuracy for most isoforms, with an average value of 0.843. SMARTCyp performs less competitively across isoforms, consistent with the limitations of fixed rule-based heuristics. GLMCyp is excluded from the fair comparison, as its reported Top-2 accuracy is obtained under an evaluation setting in which the evaluation set includes both training and test instances. Although the average Top-2 accuracy of ATTNSOM is comparable to that of XenoSite, ATTNSOM attains the best performance on a substantially larger number of CYP isoforms (6 out of 9), indicating more consistent cross-isoform generalization. In contrast, performance gains are less pronounced for CYP2D6 and CYP3A4. Consistent with Figure~\ref{fig:heatmap}, these isoforms appear far from others in the dendrogram, indicating limited shared metabolic patterns. Consequently, they benefit less from cross-isoform information sharing.

\input{tables/result_mcc}

\paragraph{\textbf{MCC Performance and Ablation}}
Table~\ref{tab:mcc} presents MCC performance for SOM prediction across nine human CYP isoforms. The Matthews correlation coefficient (MCC) evaluates atom-level binary discrimination by jointly accounting for false positives and false negatives under severe class imbalance. The full ATTNSOM model achieves the highest MCC for every CYP isoform, resulting in an average MCC of 0.710 across isoforms. Consistent improvements over all ablated variants are observed for each isoform, demonstrating the robustness and effectiveness of the proposed architecture in capturing cross-isoform metabolic patterns.

Ablation studies examine the roles and interdependence of the main architectural components. Removing the attention mechanism (ATTNSOM w/o attn.) results in a clear reduction in average MCC (0.670), while removing feature-wise linear modulation (FiLM) conditioning (ATTNSOM w/o FiLM) also leads to degraded performance (0.676). Interestingly, removing both attention and FiLM yields slightly higher performance than removing attention alone. This result indicates that FiLM conditioning is most effective when atom representations are aligned with CYP-related information through the attention mechanism. When attention is absent, FiLM lacks sufficient isoform-aligned contextual signals to guide meaningful feature modulation, providing limited benefit over a simpler baseline. In contrast, removing both modules yields a more stable representation by directly combining atom-level features with CYP isoform embeddings. These findings suggest that attention and FiLM are complementary components whose joint application is crucial for achieving optimal performance.

Replacing the molecular encoder with alternative graph backbones also leads to substantial performance degradation. Variants using Chemprop~\citep{heid2023chemprop}, GIN~\citep{xu2018powerful}, GCN~\citep{kipf2016semi}, or GAT~\citep{velickovic2017graph}  encoders exhibit consistently lower MCC values, with average scores ranging from 0.343 to 0.606, indicating that the proposed ATTNSOM encoder is better suited for capturing the structural determinants of molecular reactivity. Finally, standalone baseline models using Chemprop, GIN, GCN, and GAT show the lowest MCC values overall, underscoring the importance of the proposed architectural design beyond the choice of molecular encoder alone. These results demonstrate that ATTNSOM achieves state-of-the-art MCC performance and both attention mechanisms and FiLM-based conditioning are critical for accurate and balanced SOM prediction across diverse CYP isoforms.

\paragraph{\textbf{Performance on the AZ-ExactSOM Dataset}}
Table~\ref{tab:az} summarizes overall performance on the AZ-ExactSOM dataset using MCC, precision, recall, and Top-$k$ accuracy, which jointly evaluate balanced atom-level classification and ranking quality. Learnable methods (ATTNSOM, Chemprop, GIN, GCN, and GAT) were evaluated using 10-fold cross-validation. The proposed ATTNSOM achieves the highest MCC and recall, indicating superior overall performance and strong sensitivity in identifying true metabolic sites. Moreover, ATTNSOM attains the best Top-2 and Top-3 accuracies, demonstrating its effectiveness in prioritizing relevant SOMs among the top-ranked predictions. Chemprop demonstrates high precision and Top-1 accuracy but lower recall and MCC, reflecting limited coverage of metabolically active sites. GIN, GCN, and GAT exhibit substantially weaker performance across all metrics, with GAT failing to produce meaningful predictions under this evaluation setting. For FAME 3 and SMARTCyp, predictions are obtained via their publicly available web servers, as these methods are not distributed as trainable models. These methods show markedly lower MCC and recall compared to learning-based approaches. Across the Zaretzki and AZ-ExactSOM datasets, ATTNSOM achieves high Top-$k$ accuracy and MCC, demonstrating the effectiveness of jointly modeling molecule-contextualized atom representations and cross-isoform relationships for atom-level SOM prediction.

\input{tables/result_az}

%% file: tables/result_top2.tex
\begin{table}[htbp]
\centering
\caption{Top-2 sites of metabolism prediction accuracy across nine CYP450 isoforms evaluated using Zaretzki dataset. Bold values indicate the best-performing result.}
\label{tab:top2}
\scalebox{0.70}{
\begin{tabular}{lcccccccccc}
\toprule
\textbf{Model} & \textbf{CYP1A2} & \textbf{CYP2A6} & \textbf{CYP2B6} & \textbf{CYP2C8} & \textbf{CYP2C9} & \textbf{CYP2C19} & \textbf{CYP2D6} & \textbf{CYP2E1} & \textbf{CYP3A4} & \textbf{Average} \\
\midrule
ATTNSOM & 0.864 & \textbf{0.905} & \textbf{0.908} & \textbf{0.888} & \textbf{0.938} & \textbf{0.922} & 0.844 & \textbf{0.868} & 0.813 & \textbf{0.871} \\
XenoSite & \textbf{0.871} & 0.857 & 0.834 & 0.887 & 0.867 & 0.890 & \textbf{0.885} & 0.835 & \textbf{0.876} & 0.870 \\
RS-Predictor & 0.834 & 0.857 & 0.821 & 0.838 & 0.845 & 0.862 & 0.859 & 0.828 & 0.823 & 0.843 \\
SMARTCyp & 0.800 & 0.860 & 0.770 & 0.830 & 0.840 & 0.860 & 0.830 & 0.840 & 0.780 & 0.821 \\
\bottomrule
\end{tabular}
}
\end{table}

%% file: tables/result_mcc.tex
\begin{table*}[htbp]
\caption{Matthews correlation coefficient (MCC) results for sites of metabolism prediction across nine CYP450 isoforms on the Zaretzki dataset. The table compares the full ATTNSOM with ablated variants that remove the cross- attention module (attn.) and/or the FiLM modulation, as well as versions using alternative graph encoders. Bold values indicate the best-performing result.
}
\label{tab:mcc}
\scalebox{0.62}{
\begin{tabular}{lcccccccccc}
\toprule
\textbf{Model} & \textbf{CYP1A2} & \textbf{CYP2A6} & \textbf{CYP2B6} & \textbf{CYP2C8} & \textbf{CYP2C9} & \textbf{CYP2C19} & \textbf{CYP2D6} & \textbf{CYP2E1} & \textbf{CYP3A4} & \textbf{Average} \\
\midrule
ATTNSOM & \textbf{0.683} & \textbf{0.781} & \textbf{0.750} & \textbf{0.793} & \textbf{0.739} & \textbf{0.784} & \textbf{0.692} & \textbf{0.741} & \textbf{0.645} & \textbf{0.710} \\
ATTNSOM w/o attn. & 0.657 & 0.729 & 0.725 & 0.729 & 0.701 & 0.736 & 0.652 & 0.677 & 0.606 & 0.670 \\
ATTNSOM w/o FiLM & 0.668 & 0.707 & 0.738 & 0.751 & 0.709 & 0.745 & 0.650 & 0.712 & 0.598 & 0.676 \\
ATTNSOM w/o FiLM w/o attn. & 0.644 & 0.740 & 0.735 & 0.750 & 0.698 & 0.754 & 0.646 & 0.712 & 0.604 & 0.674 \\
\midrule
ATTNSOM w/ Chemprop & 0.580 & 0.652 & 0.665 & 0.697 & 0.614 & 0.697 & 0.572 & 0.611 & 0.551 & 0.606 \\
ATTNSOM w/ GIN & 0.556 & 0.647 & 0.671 & 0.671 & 0.587 & 0.679 & 0.577 & 0.566 & 0.542 & 0.591 \\
ATTNSOM w/ GCN & 0.475 & 0.527 & 0.572 & 0.564 & 0.549 & 0.490 & 0.503 & 0.477 & 0.473 & 0.499 \\
ATTNSOM w/ GAT & 0.344 & 0.251 & 0.375 & 0.402 & 0.370 & 0.453 & 0.298 & 0.273 & 0.312 & 0.343 \\

\midrule
Chemprop & 0.535 & 0.606 & 0.666 & 0.605 & 0.568 & 0.622 & 0.526 & 0.563 & 0.521 & 0.560 \\
GIN & 0.536 & 0.525 & 0.554 & 0.596 & 0.522 & 0.632 & 0.496 & 0.461 & 0.463 & 0.517 \\
GCN & 0.470 & 0.520 & 0.571 & 0.575 & 0.534 & 0.474 & 0.480 & 0.429 & 0.447 & 0.484 \\
GAT & 0.316 & 0.274 & 0.411 & 0.351 & 0.333 & 0.409 & 0.333 & 0.232 & 0.284 & 0.327 \\

\bottomrule
\end{tabular}}
\end{table*}

%% file: tables/result_az.tex
\begin{table}[t]
\centering
\caption{Overall metrics on the AZ-ExactSOM data set, including MCC, precision, recall, and Top-\(k\) accuracy. Bold values indicate the best-performing result.
}
\label{tab:az}
\scalebox{0.9}{
\begin{tabular}{lcccccc}
\toprule
Run & MCC & Precision & Recall & TOP1 & TOP2 & TOP3 \\
\midrule
ATTNSOM & \textbf{0.517} & 0.521 & \textbf{0.595} & 0.639 & \textbf{0.797} & \textbf{0.872} \\
FAME 3 & 0.197 & 0.415 & 0.119 & 0.432 & 0.663 & 0.790 \\
SMARTCyp & 0.250 & 0.233 & 0.414 & 0.379 & 0.494 & 0.537 \\
Chemprop & 0.480 & \textbf{0.526} & 0.512 & \textbf{0.640} & 0.778 & 0.832 \\

GIN & 0.371 & 0.393 & 0.446 & 0.494 & 0.629 & 0.714 \\
GCN & 0.258 & 0.311 & 0.268 & 0.462 & 0.603 & 0.657 \\
GAT & 0.000 & 0.000 & 0.000 & 0.213 & 0.328 & 0.400 \\


\bottomrule
\end{tabular}}
\end{table}
\vspace{-3mm}

%% file: 4_analysis.tex
\section{Analysis}

\begin{figure*}[b]
  \centering
  \includegraphics[width=1\textwidth, alt={Set of molecular diagrams for ivacaftor showing experimentally annotated sites of metabolism in the left panel and atom-level SOM predictions from four methods in subsequent panels, with predicted atoms highlighted on the molecular structure.}]{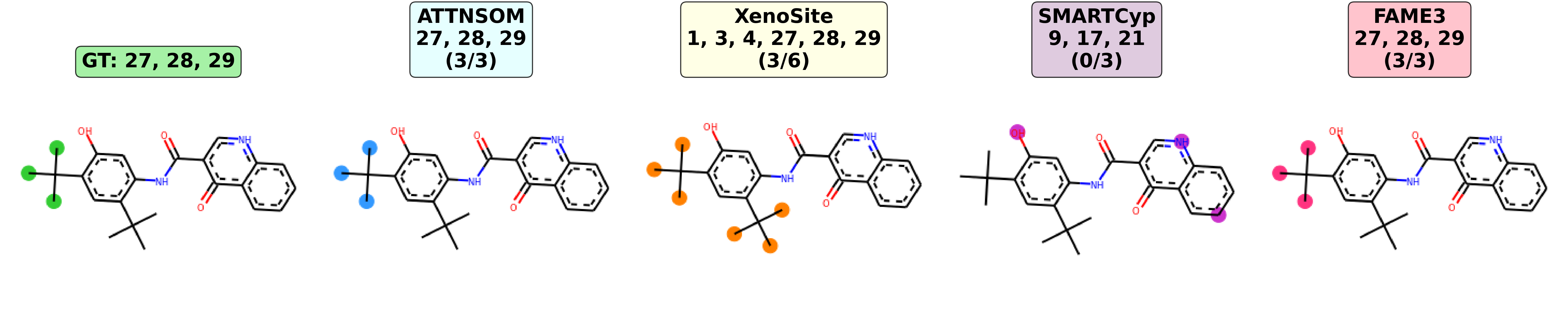}
  \caption{SOM predictions for ivacaftor illustrating atom-level localization performance across different methods. Experimentally annotated SOM atoms are shown in the left panel (GT), followed by predictions from ATTNSOM, XenoSite, SMARTCyp, and FAME 3. ATTNSOM accurately localizes all experimentally validated SOM positions, demonstrating precise atom-level discrimination and strong generalization.}
  \label{fig:ivacaftor}
\end{figure*}

\begin{figure*}[htbp]
  \centering
  \includegraphics[width=0.9\textwidth, alt={An attention heatmap with numerical attention values for three atoms across multiple CYP isoforms.}]{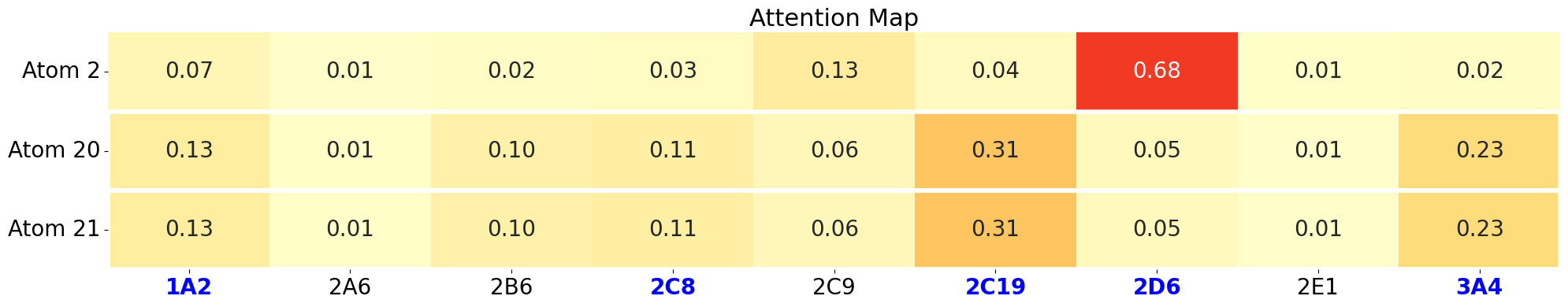}

  \vspace{0.8em}

  \includegraphics[width=0.9\textwidth, alt={The panels display molecular structures highlighting ground-truth and predicted atom-level sites of metabolism for the same molecule across five CYP isoforms.}]{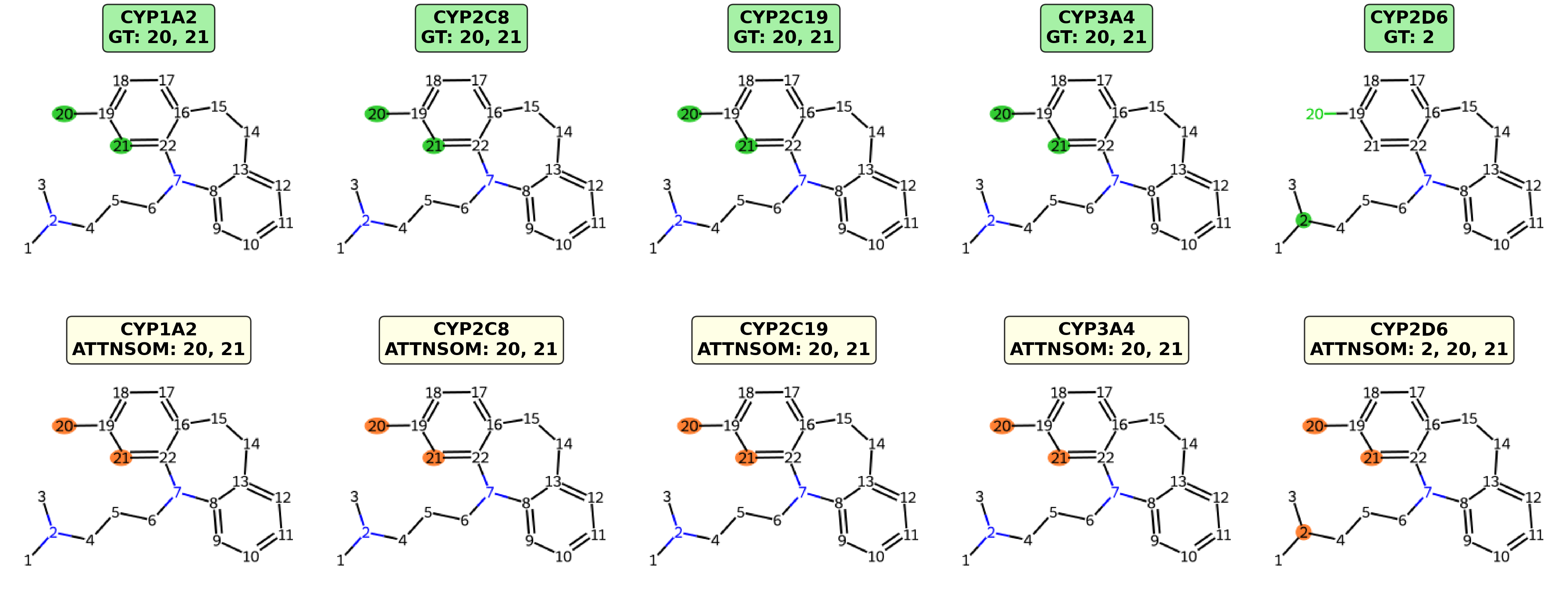}

  \caption{Attention maps and SOM predictions for a single molecule across five CYP isoforms. ATTNSOM captures shared metabolic patterns by distributing attention for atoms~20 and~21 across related CYPs, while producing sharp isoform-specific attention for CYP2D6 at atom~2. Conservative co-prediction of atoms~20/21 for CYP2D6 reflects a trade-off of cross-isoform relational modeling.} 
  \vspace{-1.0em}
  \label{fig:failure}
\end{figure*}

To provide deeper insight into the behavior of the proposed model beyond quantitative performance metrics, we conduct qualitative analyses on representative compounds as well as challenging prediction scenarios. These analyses are designed to investigate how ATTNSOM achieves strong generalization, identifies dominant metabolic sites at atomic resolution, and captures cross-isoform relationships. Through systematic comparisons, we elucidate both the strengths and the inherent trade-offs of cross-isoform SOM modeling.

\paragraph{\textbf{Ivacaftor as a Test of Generalization}}
To further evaluate whether computational SOM prediction can recover metabolically actionable sites relevant to practical drug optimization, we examine the ivacaftor–deutivacaftor pair as a representative case study. Ivacaftor contains two chemically similar tert-butyl substituents, yet oxidative metabolism has been experimentally shown to occur preferentially at only one of these sites~\citep{di2023deuterium}. The metabolic relevance of this position is further supported by the development of deutivacaftor, where selective deuteration of the corresponding tert-butyl methyl group reduced CYP3A4-mediated oxidation. We therefore define this experimentally validated position as the ground-truth SOM in our analysis. 

Figure~\ref{fig:ivacaftor} presents a qualitative comparison of SOM predictions for ivacaftor across different computational methods. The ATTNSOM prediction is obtained using weights trained on the Zaretzki dataset, which does not include ivacaftor, whereas predictions from the other methods are generated via their respective publicly available web servers. ATTNSOM accurately identifies all experimentally validated SOM positions. The predicted sites are obtained by averaging inference outputs from models trained across the 10 folds of cross-validation. By comparison, XenoSite predicts a broader set of candidate sites spanning both tert-butyl groups, reflecting limited specificity in distinguishing among chemically similar methyl substituents. SMARTCyp, which relies on predefined heuristics based on chemical reactivity and accessibility, does not recover the experimentally observed SOM. FAME 3 successfully highlights the reactive tert-butyl group. This case study underscores the practical value of precise SOM prediction for drug optimization. Accurate localization enables targeted chemical modifications, as exemplified by the improved pharmacokinetic properties of deutivacaftor.

\paragraph{\textbf{Visualization of SOM Prediction}}

Figure~\ref{fig:failure} visualizes the atom–isoform attention maps and corresponding SOM predictions with five CYP isoforms (CYP1A2, CYP2C8, CYP2C19, CYP3A4, and CYP2D6). The ground-truth SOMs (atoms 20 and 21) for CYP1A2, CYP2C8, CYP2C19, and CYP3A4 receive attention from the respective isoforms, reflecting shared metabolic preferences among related isoforms. This means that attention map effectively learns cross-isoform relationships.

The molecular panels below the attention maps show the final SOM predictions generated by the prediction head based on relational information derived from cross-attention. The model correctly predicts atoms 20, 21 as the SOMs for CYP1A2, CYP2C8, CYP2C19, and CYP3A4, and also accurately identifies atom 2 as the true SOM for CYP2D6. However, in the CYP2D6 column of the attention map, attention to atoms 20 and 21 is not fully suppressed, resulting in a conservative co-prediction together with atom 2. In contrast, atom 2 also receives a moderate attention weight in the CYP1A2 colunm, yet the prediction head selects only atoms 20, 21 as the SOMs. This behavior reflects the sharply localized CYP2D6-specific attention in the atom 2 row, while atoms 20 and 21 exhibit more diffusely distributed attention across multiple isoforms. We suggest that the reason is that strongly isoform-specific attention leads the model to suppress signals from other isoforms, whereas broadly distributed attention causes the model to interpret the site as a globally shared metabolic hotspot and respond even to relatively weak signals.

These observations highlight both the strengths and limitations of relational modeling in ATTNSOM. While the model effectively captures cross-isoform relationships and benefits from them for accurate predictions, it can also be influenced by broadly shared attention signals in cases where isoform-specific attention alone should be prioritized.  This can occasionally lead to the over-prediction of globally reactive sites, which represents a natural trade-off inherent in cross-isoform relational modeling.

\paragraph{\textbf{Learned Cross-Isoform Relationships}}

Figure~\ref{fig:Pred_Heatmap} shows the SOM pattern similarity predicted by ATTNSOM and this pattern reveals several biologically meaningful trends. In this dendrogram, CYP2A6 and CYP2E1 emerge as a closely related pair, reflecting their preference for relatively small and low-complexity substrates~\citep{zanger2013cytochrome}. Similarly, CYP2C19 and CYP2C8 show strong similarity, consistent with their classification within the CYP2C family and their overlapping substrate scope and regioselectivity characteristics~\citep{zanger2013cytochrome}. In contrast, CYP2D6 and CYP3A4 exhibit SOM patterns that are more distinct from those of other CYP isoforms. CYP2D6 preferentially metabolizes basic, positively charged substrates, and CYP3A4 accommodates a wide range of bulky and conformationally flexible compounds subject to distinct accessibility constraints~\citep{zanger2013cytochrome}. {These properties of CYP2D6 and CYP3A4 account for their separation in the similarity landscape.

The ATTNSOM-based similarity (Figure~\ref{fig:Pred_Heatmap}) does not perfectly but closely matches the dataset-based similarity (Figure~\ref{fig:heatmap}), supported by a high cophenetic correlation between the two dendrograms(r = 0.803;~\citep{876324aa-47bd-32ef-8a3b-7042924a2889}). This can be attributed to the fact that ATTNSOM primarily learns cross-isoform SOM patterns from the data, with supervised attention as an auxiliary guide rather than memorizing annotation patterns. For example, while the dataset-level similarity space suggests a closer similarity between CYP1A2 and CYP2C8/CYP2C19, the predicted similarity space places CYP2B6 closer to CYP2C19 and CYP2C8. This difference reflects overlapping metabolic behavior for a subset of substrates emphasized by the learned representations. Indeed, the metabolic behavior of CYP2B6 and CYP2C enzymes is driven by their shared preference for bulky, drug-like compounds, in contrast to the smaller, planar aromatic substrates favored by CYP1A2~\citep{zanger2013cytochrome}. This highlights ATTNSOM's ability to move beyond the limitations of imperfect data annotations and capture functionally meaningful, substrate-driven metabolic relationships.

\begin{figure}[t]
  \centering
  \includegraphics[width=0.8\textwidth, alt={Heatmap displaying pairwise SOM pattern similarity values predicted by ATTNSOM across CYP isoforms, with a hierarchical clustering dendrogram above the matrix grouping isoforms based on similarity.}]{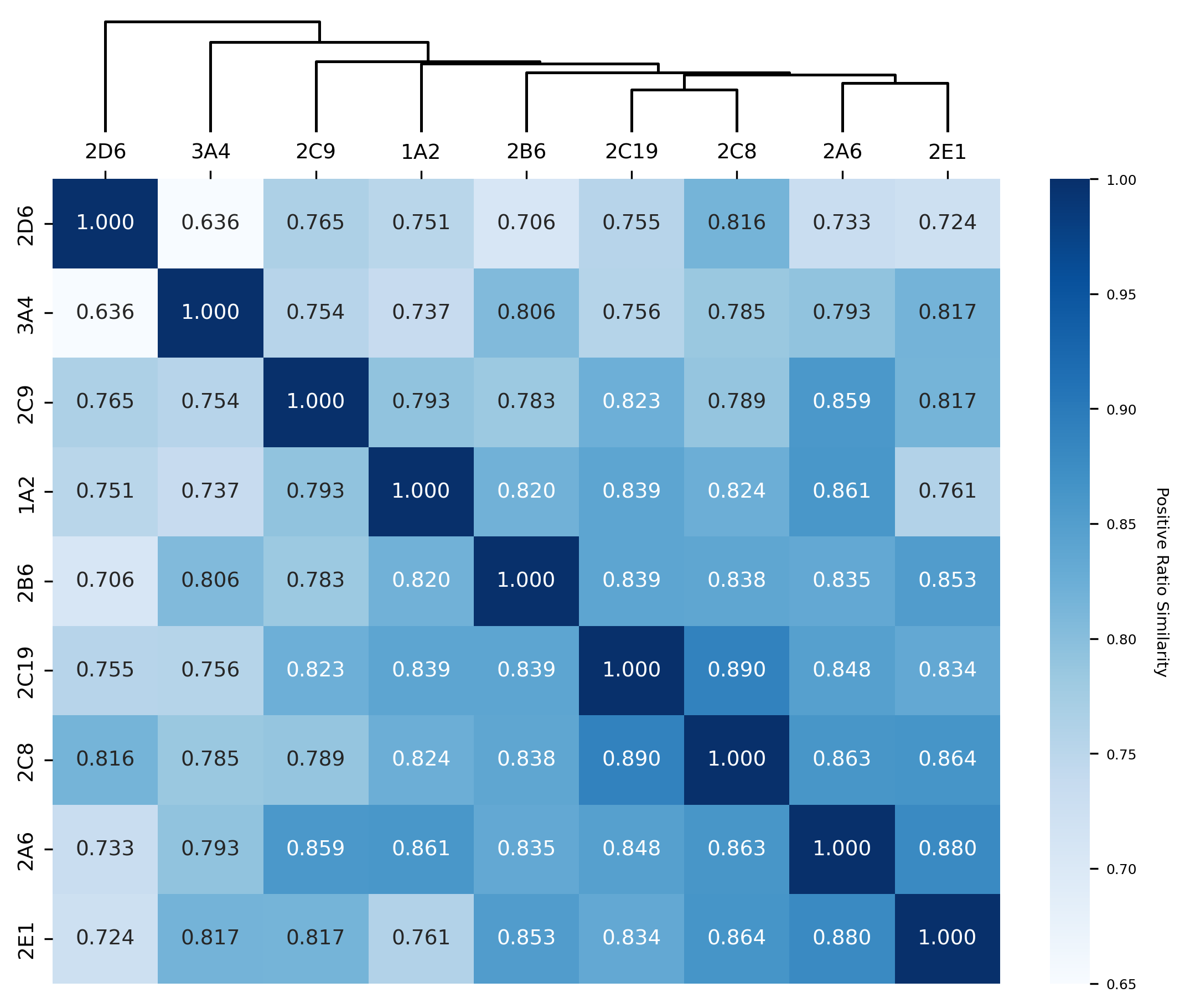}
  \caption{Hierarchical clustering and heatmap of pairwise SOM pattern similarity predicted by ATTNSOM across CYP isoforms.}
  \label{fig:Pred_Heatmap}
\end{figure}

%% file: 5_conclusion.tex
\section{Discussion}
Extensive quantitative evaluations show that ATTNSOM performs competitively across multiple CYP isoforms, achieving strong Top-$k$ accuracy together with improved Matthews correlation coefficient (MCC). This combination indicates that ATTNSOM effectively ranks reactive sites while maintaining balanced atom-level discrimination under severe class imbalance. Qualitative analyses further demonstrate ATTNSOM’s generalization ability and its modeling of cross-isoform relationships. These analyses show that the model adapts its predictions to isoform-specific preferences and captures biologically meaningful cross-isoform relationships, while remaining constrained by the coverage of available annotations.

However, our analysis reveals a limitation of ATTNSOM when atoms exhibit strong isoform-agnostic reactivity across multiple CYPs. In such cases, cross-isoform attention assigns non-negligible attribution to these atoms even for isoforms with otherwise sharp metabolic preferences, leading to conservative co-prediction. This behavior arises from the atom-centric attention formulation, in which atom representations query multiple CYP isoforms and aggregate shared metabolic signals. While this design improves robustness and generalization by capturing globally plausible sites, it can reduce isoform-specific localization precision by over-predicting sites that are broadly reactive but not dominant for a given isoform. Future work will focus on addressing this trade-off by modeling adaptive attention mechanisms that dynamically regulate information sharing based on the degree of metabolic overlap among target isoforms. Incorporating additional biochemical or structural priors may further improve specificity for closely related CYP enzymes.

\section{Conclusion}

We introduce ATTNSOM, a cross-isoform relation learning framework for atom-level prediction of CYP-mediated site-of-metabolism. The framework addresses key limitations in existing computational approaches by integrating three essential components: (1) a shared graph encoder that captures intrinsic chemical reactivity independent of enzyme context, (2) FiLM-based molecule-specific modulation of atom representations to incorporate molecular context and allow features to vary across different molecular environments, and (3) cross-attention mechanisms that model shared metabolic patterns and relationships across CYP isoforms. Rather than treating isoforms independently, ATTNSOM exploits correlated metabolic patterns to improve SOM prediction accuracy and discrimination reliability across multiple CYP enzymes.

%% file: 6_extra.tex
\section{Funding}
This work was supported by the National Research Foundation of Korea (NRF) grant funded by the Korea government, specifically the Ministry of Science and ICT (MSIT) and the Ministry of Education (MOE) (No. RS-2025-16652968). This research was also supported by the National Research Foundation of Korea (NRF) under grant No. NRF-2023R1A2C3004176, and by the grant HR20C002103. In addition, this work was supported by a grant of the Korea Machine Learning Ledger Orchestration for Drug Discovery Project (K-MELLODDY), funded by the Ministry of Health and Welfare and the Ministry of Science and ICT, Republic of Korea (grant No. RS-2024-00462471).